\documentclass[aps,prl,twocolumn,showpacs,amsmath,amssymb]{revtex4}

\usepackage{wrapfig}
\usepackage{sidecap}
\usepackage{graphicx,pslatex}
\usepackage{dcolumn}
\usepackage{bm}

\begin{document}

\title{Phase Resolved Surface Plasmon Interferometry of Graphene}

\author{Justin A. Gerber} 
\thanks{These authors contributed equally to this work.}
\author{Samuel Berweger$^*$} 
\altaffiliation{Current address: National Institute of Standards and Technology, Boulder, CO, 80305}
\author{Brian T. O'Callahan}
\author{Markus B. Raschke}
\email{markus.raschke@colorado.edu}
\affiliation{Department of Physics, Department of Chemistry, and JILA, University of Colorado, Boulder, CO 80309, USA}

\date{\today}

\begin{abstract}
The surface plasmon polaritons (SPP) of graphene reflect the microscopic spatial variations of underlying electronic structure and dynamics. 
Access to this information requires probing the full SPP response function. 
We image the graphene SPP phase and amplitude by combining scanning probe tip coupled surface plasmon interferometry with phase resolved near-field signal detection. 
We show that a simple analytical cavity model can self-consistently describe the phase and amplitude response both for edge, grain boundary, and defect SPP reflection and scattering. 
The derived complex SPP wavevector, damping, and carrier mobility agree with the results from more complex models.
This phase information opens a new degree of freedom for spatial and spectral graphene SPP tuning and modulation for opto-electronics applications. 
\end{abstract}

\maketitle

The light-matter interaction of graphene is distinct from that of other forms of matter due to its unique electronic band structure. 
The high quantum yield has already enabled a range of opto-electronics and photonic applications based on single particle excitations.
However, even more unusual are the collective particle excitations in the form of Dirac plasmons, typically in the infrared spectral range, with their properties controllable by electric field gating, doping, or multilayer stacking \cite{hwang07,jablan09,ju11,yan12,roldan13}.
In the long wavelength limit the unique properties of massless Dirac fermions lead to a very large reduction of surface plasmon polariton (SPP) wavelength compared to the free space excitation wavelength: $\lambda_{\rm SPP} / \lambda_0 \simeq 2 \alpha E_F / (\epsilon \hbar \omega) \sim \alpha$ \cite{grigorenko12}, with the fine structure constant $\alpha$.
The short SPP wavelength gives rise to a strong spatial confinement, but the momentum mismatch due to the associated large in plane wave vector concomitantly requires high k-vector field components for the SPP excitation. 
That coupling can be achieved through the near-field of the nanometer scale apex of a scanning probe tip \cite{hecht96,ren11,fei11}.
Using the tip to excite and subsequently scatter the Dirac plasmons polarization into 
detectable far-field
radiation, the expected deep-sub-wavelength 
SPP
standing wave from boundary reflections at mid-IR frequencies could be imaged using scattering-type scanning near-field optical microscopy (\emph{s}-SNOM) and described theoretically \cite{fei12,chen12,fei13,chen13}.
\begin{figure}[!htb]
\includegraphics[width=\columnwidth]{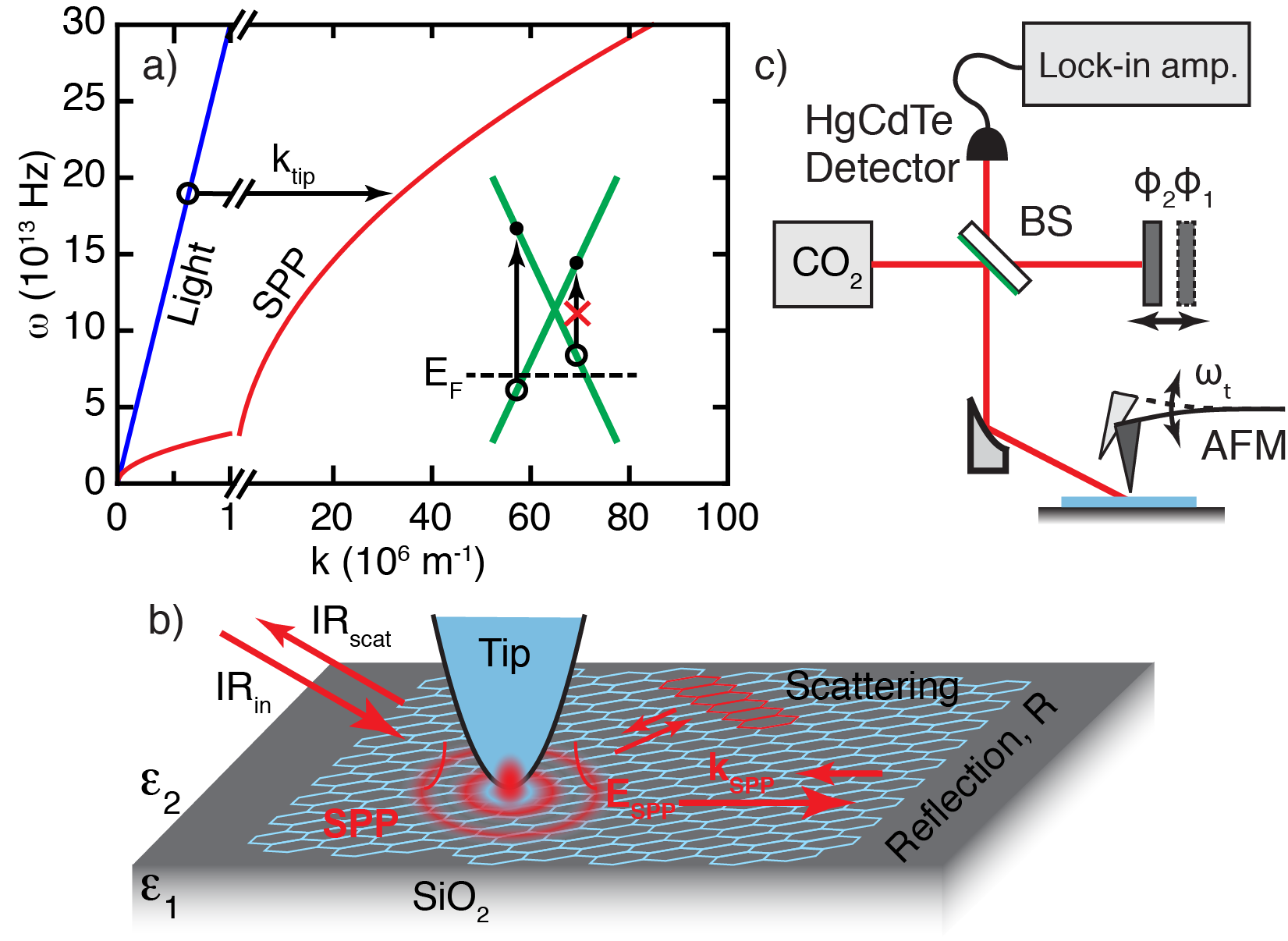}
\caption{(color online) a) Graphene SPP dispersion relation for $E_F$~=~0.4~eV.
Inset: Pauli blocking arising from doping-induced Fermi level shift.
b) Illustration of tip-induced SPP excitation and subsequent interference due to emission of scattered and reflected SPP waves.
c) Schematic of the experimental setup.
}
\label{setup} 
\end{figure}

The plasmon wavelength, its damping, and other spatial details directly relate to the local electronic structure, which is determined by doping or strain \cite{hwang07,jablan09,ni08}, the number of layers and their stacking order \cite{roldan13,sensarama10,mak10}, and is further modified by atomic scale discontinuities at edges, grain boundaries, and defects.
With the exquisite sensitivity of the spatial plasmon response to these parameters, near-field plasmon interferometry can serve as a sensitive probe for the electronic structure and its spatial inhomogeneities, which are difficult to access by other techniques. 
 
However, near-field imaging experiments have so far only analyzed the amplitude of the optical response \cite{fei12,chen12,fei13,chen13}.
The full characterization of \emph{s}-SNOM amplitude and phase response is not only desirable in general \cite{berweger13,taubner04}.
For graphene in particular, it it not only necessary to provide a complete understanding of its optical response function, but can provide complementary self-consistent information to validate proposed models.

Here, we provide near-field interferometric full {\em phase} and {\em amplitude} resolved spatial imaging of graphene plasmons reflected and scattered by external (edges) and internal boundaries (folds and grain boundaries), as well as defects, of single and multilayer graphene. 
That independent measurement and control of the full spatial optical response function of the plasmon field provides additional and direct information about differences in phase behavior of the plasmon interaction with the reflecting internal and external boundaries.
Moreover, the knowledge of the SPP response function in terms of amplitude and phase allows us to develop a simple cavity model that provides a full, self-consistent, and intuitive description of the essential optical physics of the graphene plasmons and complements the more complex numerical electrodynamic theories. 
We derive plasmon wavevector, damping, and carrier mobility, with values in agreement with theory.

In the spectral region $\hbar \omega<2E_F$, where Pauli blocking occurs, graphene exhibits a Drude-type behavior \cite{li08} and the SPP wavevector is given in turn by the local Fermi level $E_F$ and free-space wavelength $\lambda_0$ by \cite{jablan09}
\begin{equation}
k_{\rm SPP}=\frac{2 \pi h^2 c^2 \epsilon_0 \kappa}{e^2 E_F \lambda_0^2},
\label{dispersion}
\end{equation}
where $\kappa$ is the average dielectric function of the embedding media $\kappa=\kappa_1+i \kappa_2=(\epsilon_1+\epsilon_2)/2$.
Fig.~\ref{setup}(a) shows the calculated dispersion relation of the graphene SPP, assuming $E_F$~=~0.4~eV and $\kappa=2.5$ (corresponding to SiO$_2$/air at $\lambda_0$~=~10.8~${\rm \mu m}$ \cite{palik}) as an example, and compared to the light line (blue).
The large in-plane momentum necessary to overcome that wavevector mismatch is provided by the evanescent near-field of the tip with apex radius $r \propto 1/k_{\rm tip}$ \cite{hecht96,ren11,fei11}. 
SPP's are thus launched, and subsequently scattered or reflected at electronic inhomogeneities in the form of, e.g., defects \cite{fei12}, edges \cite{fei12,chen12}, or 
other structural discontinuities \cite{fei13,chen13} as illustrated in Fig.~\ref{setup}(b).
These reflected waves, after propagating back to the tip, interfere with other local near-field signal contributions and are scattered by the tip into the far-field where they are detected \cite{chen12,fei12}. 

In the experiment as shown in Fig.~\ref{setup}(c) a $^{13}$CO$_2$ laser (Access Laser, $\lambda$~=~10.8 $\mu$m) is focused onto the tip of an atomic force microscope (AFM, Anasys Instruments) operating in tapping mode using an off-axis parabolic mirror (NA~=~0.35,  P $\sim$ 5~mW).
The tip-scattered near-field $E_{\rm nf}$ is homodyne amplified at the HgCdTe detector (Kolmar Technologies) with the reference field $E_{\rm ref}$ from the other arm of the Michelson interferometer.
The far-field background is suppressed by lock-in demodulation (Zurich Instruments) at the third harmonic of the cantilever frequency \cite{keilmann04}.
In order to further suppress amplification of the near-field by the self-homodyne background $E_{\rm bg}$ with uncontrolled phase \cite{aubert03} a strong reference field $E_{\rm ref}/E_{\rm bg}\ge 10$ is used \cite{jones09}.
By collecting raster-scanned images at two orthogonal reference phases, the full complex valued tip-scattered near-field $\tilde{A} = A e^{i\Phi}$ can be determined with low error \cite{berweger13}.
Mechanically exfoliated graphene \cite{novoselov05} on SiO$_2$ was obtained commercially (Graphene Industries).

\begin{figure}[tb]
\includegraphics[width=0.95\columnwidth]{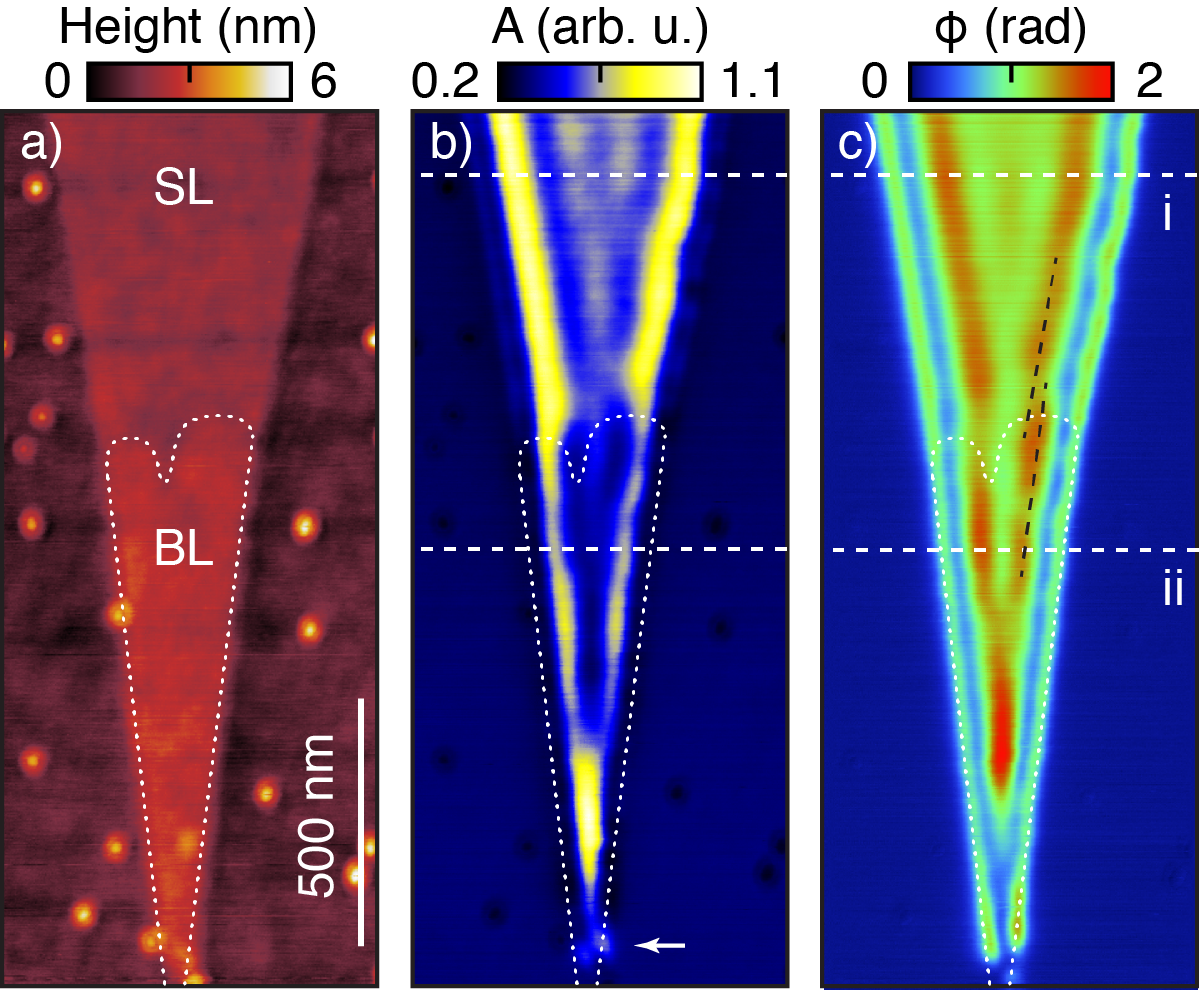}
\caption{
(color online) AFM topography (a), near-field amplitude $A=|\tilde{A}|$ (b), and near-field phase $\phi=\rm{Arg}(\tilde{A})$ (c) of a graphene wedge.
The transition from single layer (SL) to bilayer (BL) graphene is indicated. 
}
\label{wedge}
\end{figure}

Fig.~\ref{wedge} shows a typical image of a high-aspect ratio graphene wedge, chosen to feature both single and bilayer regions as indicated in the topography (a), with \emph{s}-SNOM amplitude $A=|\tilde{A}|$ (b) and phase $\phi=\rm{Arg}(\tilde{A})$ (c).
The detected tip scattered light is a superposition of the intrinsic sample optical response that is expected to be largely independent of tip position, and the SPP waves whose properties will be a function of the local environment or geometry.
The amplitude thus exhibits a standing wave pattern from the local interference of SPP reflection from both edges of the wedge, with a maximum seen at a distance of $\lambda/4$ in good agreement with previous studies \cite{chen12,fei12}.
Phase and amplitude standing waves both exhibit a periodicity of $\lambda/2$ and differ by $\sim$90$^\circ$.
A distinct feature is the phase maximum pinned to the edge.
The bilayer region (dotted line) at the tapering end of the wedge is characterized by a decreased amplitude, in addition to reduced $\lambda_{\rm SPP}$ as seen by a shifting of the maxima in both the amplitude and phase (black dashed lines) closer to the edge of the wedge.

\begin{figure}[!htb]
\includegraphics[width=0.95\columnwidth]{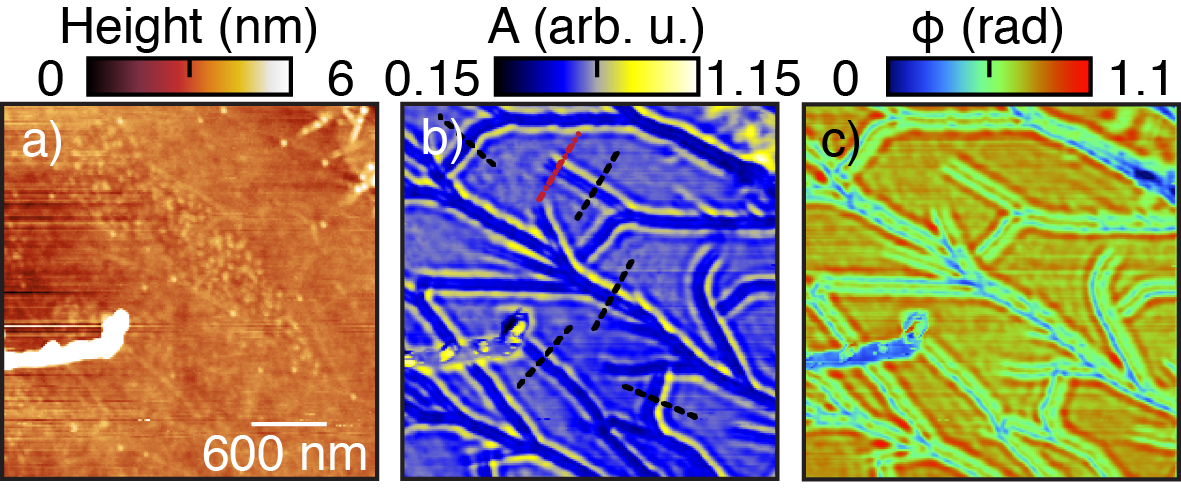}
\caption{
(color online) AFM topography (a), near-field amplitude (b), and near-field phase (c) of a region of single-layer graphene with a high concentration of grain boundaries and folds.
}
\label{boundaries}
\end{figure}

Fig.~\ref{boundaries} shows the corresponding behavior of SPP's at grain boundaries and folds, with AFM topography (a), SPP amplitude (b), and phase (c) of monolayer graphene with a high density of both kinds of linear defects.
As seen in both $A$ and $\phi$, plasmon reflection and standing wave behavior are observed and are qualitatively similar to the external boundaries.
The spatial signal variations along the boundaries and within the graphene domains are highly reproducible, including variations in SPP wavelength and amplitude across boundaries.

\begin{figure}[tb]
\includegraphics[width=.95\columnwidth]{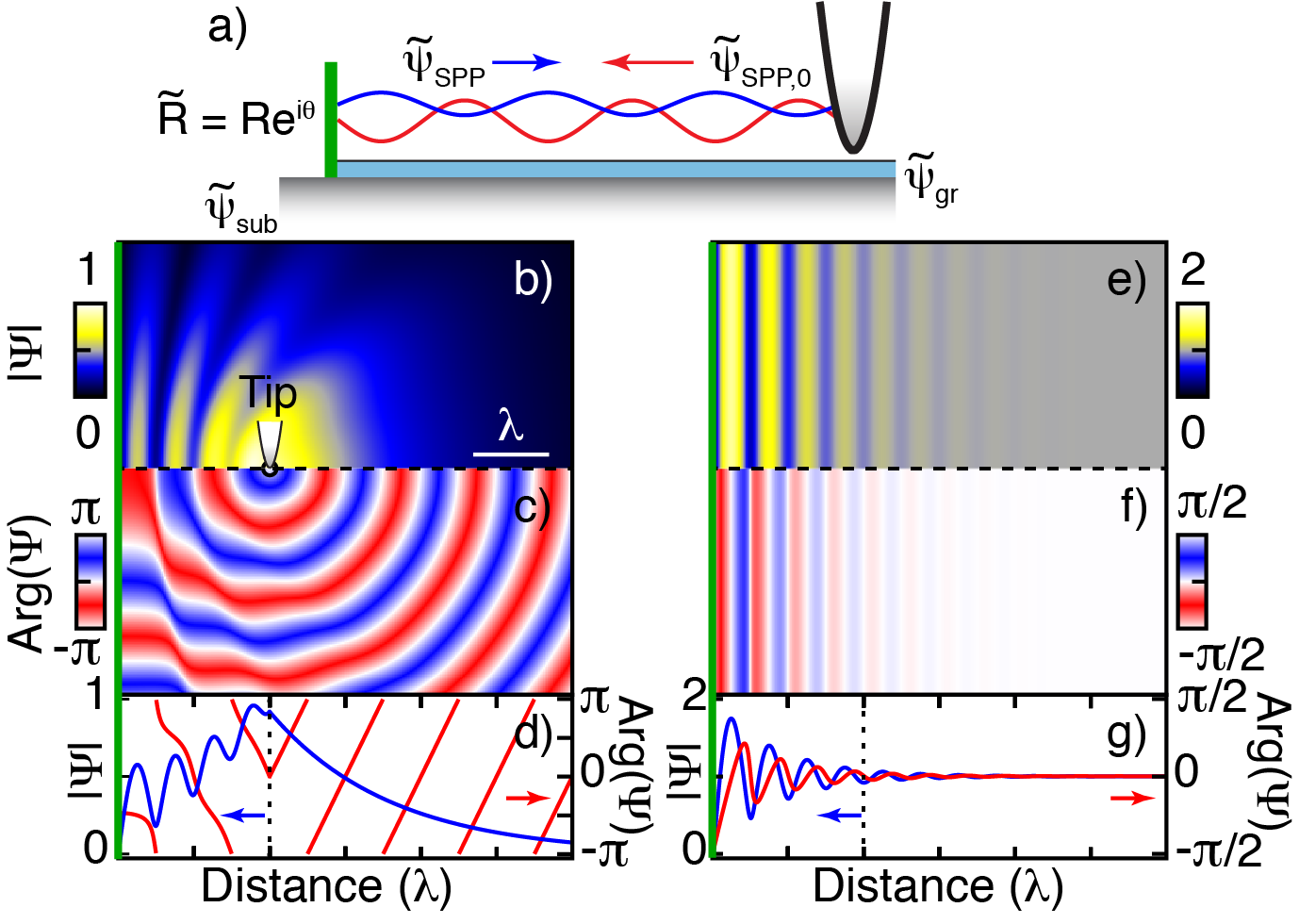}
\caption{
(color online) Illustration of graphene SPP cavity model (a). 
Calculated SPP distribution of amplitude (b), phase (c), and corresponding line cut along dashed line (d), with local tip excitation and reflection from a boundary at the left edge. 
Resulting spatial standing wave SPP map of {\em s}-SNOM amplitude (e), phase (f), and line cut (g) when scanning the tip.}
\label{model}
\end{figure}

Previous models described the measured SPP interference images to good agreement \cite{chen12,fei12,chen13,fei13}.
However, they relied on complex numerical approaches, and did not address the phase.
Here we show that the SPP oscillations can be described in both amplitude and phase simultaneously using a simple phenomenological cavity model with no independent parameters as shown schematically in Fig.~\ref{model}(a).
The tip-scattered near-field response of graphene is the sum of a non-resonant dielectric contrast contribution $\tilde{\psi}_{\rm gr}$, a resonant local tip-induced SPP term $\tilde{\psi}_{\rm SPP,0}$, and the reflected SPP fields $\tilde{\psi}_{\rm SPP,i}$, as
\begin{equation}
 \Psi_{\rm gr} =\tilde{\psi}_{\rm gr} +\tilde{\psi}_{\rm SPP,0}+\sum_{i}\tilde{\psi}_{\rm SPP,i},
 \end{equation}
each with respective amplitude and relative phase as illustrated in Fig.~\ref{model}(a).
In addition we consider a SiO$_2$ substrate near-field response $\tilde{\psi}_{\rm sub}$.
We describe $\tilde{\psi}_{\rm SPP,i}=\tilde{R_i} \times \tilde{\psi}_{\rm SPP,0}\,\,{\rm exp}\{{-2{\rm Re}(k_{\rm SPP})r_i(\gamma+i )}\}$ with decay constant $\gamma$, distance between tip and reflection $r_i$, and complex valued scattering coefficient $\tilde{R_i}$.

In order to account for the finite size of the tip apex generating a spatially averaged near-field response we use a weighting function $\Theta$ convolved with the spatially varying optical response $\Psi$ to give the simulated \emph{s}-SNOM amplitude $\tilde{A}(r)=\Psi(r') \ast\Theta(r-r')$.
As discussed below we find that treating $\Theta$ as Gaussian to approximate the evanescent nature of the tip near-field, peaked at $r$, with a width of 11 nm to model the tip radius, provides good agreement with the experimental results. 

Fig.~\ref{model} shows the calculated spatial distribution of SPP amplitude (b) and phase (c) for a stationary point source (tip) located at a fixed distance from a reflecting straight boundary located on the left edge (green line). 
For all calculations we use $\tilde{\psi}_{\rm gr}=0$, $\tilde{\psi}_{\rm SPP,0}=1$, $\tilde{R}=-1$, and $\gamma = 0.1$. 
The line cut (d) taken along the dashed line separating (b) and (c) shows the propagation of the SPP away from the tip and its subsequent perturbation by the reflected SPP.
Scanning the tip then results in a parallel standing wave pattern in spatial \emph{s}-SNOM amplitude (e) and phase (f) as seen experimentally.
From the line cut (g) we see a standing wave period of $\lambda/2$ for both phase and amplitude in agreement with experiment (for extended discussion of model and origin of observed phenomena see supplement).

\begin{figure}
\includegraphics[width=.95\columnwidth]{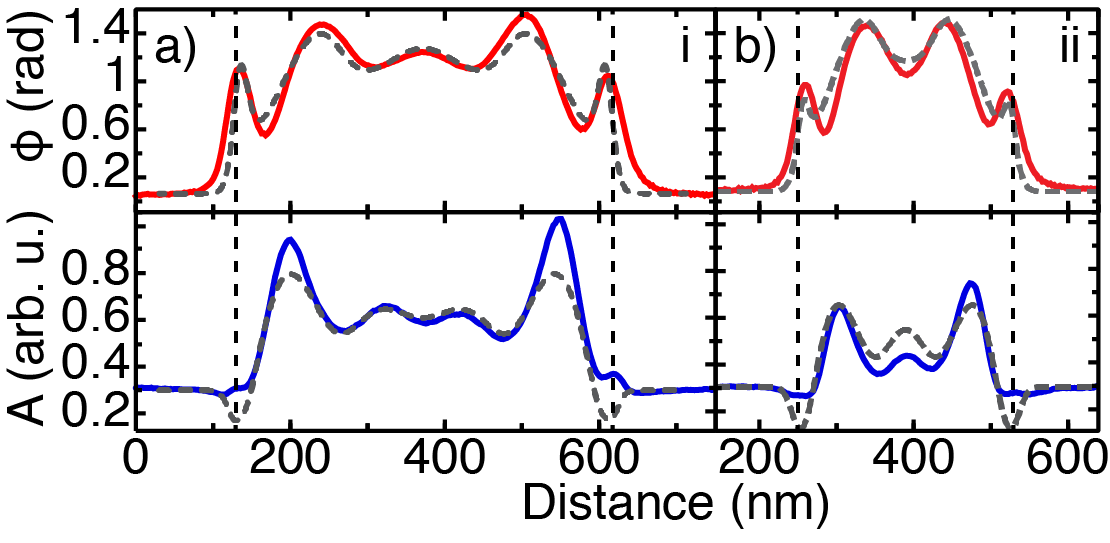}
\caption{
(color online) Line cuts of phase (red) and amplitude (blue) taken along the dashed lines in Fig.~\ref{wedge} with (i) shown in (a) and (ii) in (b).
Dashed lines are fits to cavity model for a single parameter set.
}
\label{cut}
\end{figure}

We first compare the measured \emph{s}-SNOM signal with our model by examining line cuts in Fig.~\ref{wedge} along the dashed lines.
Fig.~\ref{cut} shows experimental phase (red) and amplitude (blue) along the dashed lines (i) and (ii) in Fig.~\ref{wedge}, with (a) and (b) corresponding to single layer and bilayer graphene, respectively.
Graphene Edges on both sides are assumed to have identical reflection and decay parameters for $\tilde{\psi}_{\rm SPP,i}$.
The dashed grey lines show the results of the model calculations, reproducing for just a single parameter set for the single and bilayer, respectively, all the main spatial features both in amplitude and phase simultaneously. 
The only exception is a larger than predicted decrease in amplitude at the graphene edge as discussed below.

The best agreement between theory and experiment is obtained for a reflection coefficient of $R \approx -1$, corresponding to a $\pi$ phase shift, with SPP wavelength of $\lambda_{\rm SPP} = (260 \pm 10)$ nm, and $\gamma = 0.25 \pm 0.04$ for the single layer (a).
For the bilayer (b) we find a shorter wavelength of $\lambda_{\rm SPP} = (190 \pm 10)$ nm as discussed below, yet the same damping and reflection coefficients.
We further find a phase difference of $\sim65^\circ$ between $\tilde{\psi}_{\rm gr}+\tilde{\psi}_{\rm SPP,0}$ and SiO$_2$ substrate response $\tilde{\psi}_{\rm sub}$ for both single layer and bilayer graphene.
This phase shift is less than the 90$^\circ$ expected between the resonant SPP and non-resonant substrate response.
However, because graphene does not entirely screen the tip-substrate interaction, signal contributions from the underlying SiO$_2$ reduce the overall phase shift.
While the phase for bilayer and single layer graphene are identical, an overall smaller amplitude of $\tilde{\psi}_{\rm gr}+\tilde{\psi}_{\rm SPP,0}$ is found for bilayer graphene.

\begin{figure}[tb]
\includegraphics[width=0.7\columnwidth]{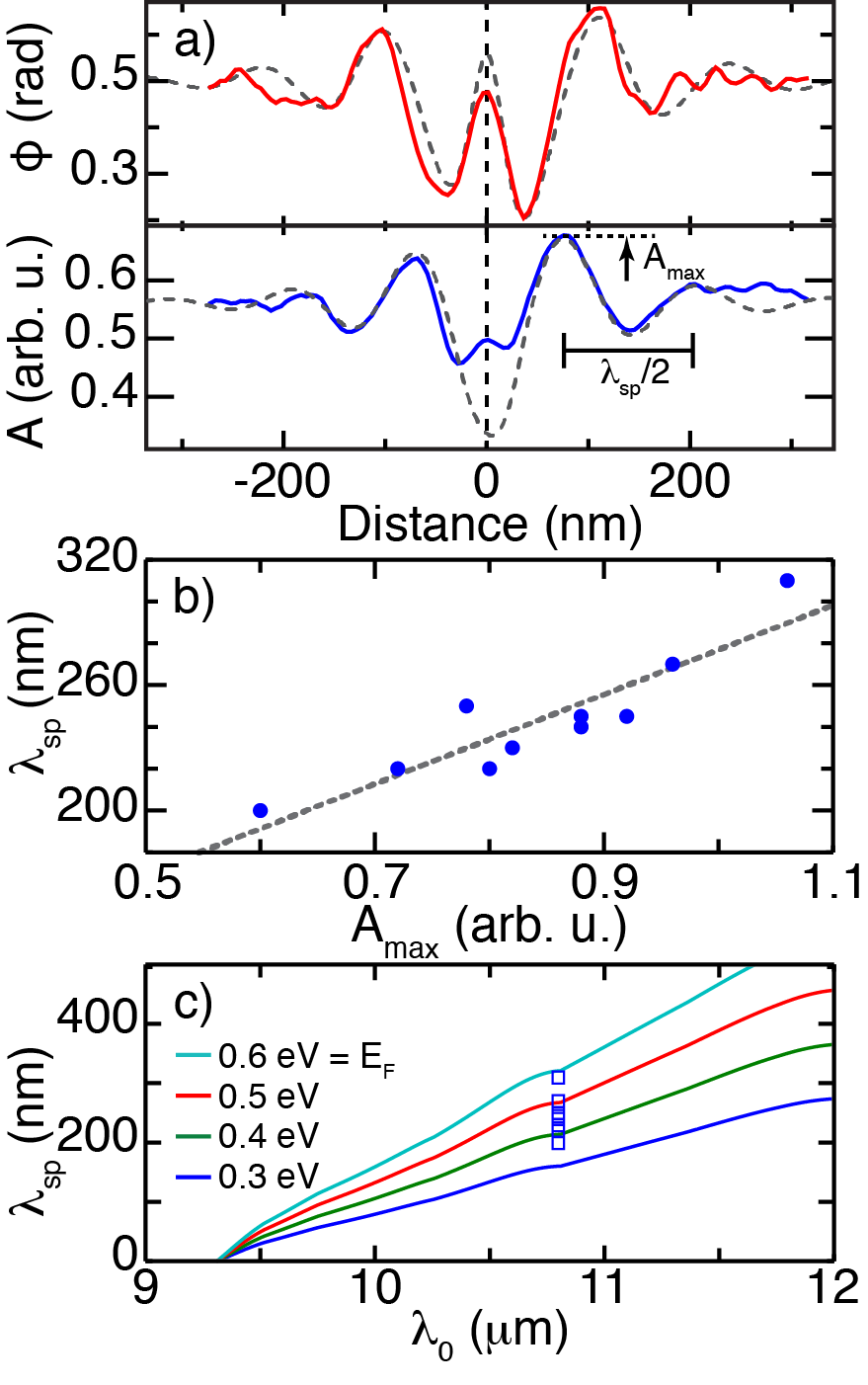}
\caption{
(color online) (a) Line cut along the red dashed line in Fig.~\ref{boundaries} showing phase (red) and amplitude (blue) along with modeled response (grey dashed).
(b) Plot of SPP wavelength vs. amplitude extracted from both sides of linear defects from Fig.~\ref{boundaries}. 
Dashed line is a linear fit.
(c) SPP dispersion relation for varying graphene Fermi levels as indicated. 
Square symbols indicate range of experimental $\lambda_{\rm SPP}$ from (b) reflecting the local variations in dopant concentration.}
\label{scatter}
\end{figure}

To model the observed spatial SPP behavior at the different internal interfaces, we examine the line cut in Fig.~\ref{boundaries}(b) (red dashed line) as an example.
Resulting phase and amplitude traces are shown in Fig.~\ref{scatter}(a) together with the result of the model (grey dashed lines).
Note that the left (right) sides of the boundary have different SPP wavelengths of $\lambda_{\rm SPP}$~=~240 nm (260 nm), as well as different reflection coefficients of $\tilde{R}$~=~0.45 (0.55).
The different wavelengths and thus wavevectors indicate a possible difference in electronic structure on either side of the linear defects in that region.
Variations in reflection coefficients can be explained by the change in SPP wavevector across the boundary, where momentum conservation facilitates transmission from the high-momentum to low-momentum side similar to material interfaces in conventional optics.

To determine and model possible local variations in electronic properties that underly the complex optical response, Fig.~\ref{scatter}(b) shows the relationship between $\lambda_{\rm SPP}$ and maximum amplitude $A_{\rm max}$ extracted from both sides of a series of representative boundaries indicated by black dashed lines in Fig.~\ref{boundaries}(b).
A clear correlation of an increase in SPP intensity $A_{\rm max}$ with an increase in $\lambda_{\rm SPP}$ is observed.
Since $A_{\rm max}$ is directly related to the reflection coefficient, this correlation is due to the reduced reflection of high-momentum SPP's.

Using Eq.~\ref{dispersion} we can relate that variation in SPP wavelength to local variations in the Fermi level.
Shown in Fig.~\ref{scatter}(c) are the calculated SPP dispersion relations of graphene on SiO$_2$ for different values of $E_F$.
The blue squares indicate the range of wavelengths from (b), thus corresponding to spatial variations in $E_F$ of up to 0.2~eV.
The Fermi level directly relates to doping concentration $n$ as $E_F=\hbar v_F k_F$, with Fermi velocity $v_F\approx 1 \times 10^6$ m/s and Fermi momentum $k_F=\sqrt{\pi n}$ \cite{castroneto09}.
Our spatial variations in Fermi level of $E_F \approx$~0.4~--~0.6~eV thus correspond to $n$ ranging from 1.2~--~2.6$\times$10$^{13}$~cm$^{-2}$.

The large variation in doping across the boundaries indicates that the reduced conductivity of boundaries \cite{tapaszto12,fei13} prevents charge equilibration between adjacent sides.
Unlike the charge carriers themselves, the SPPs as collective excitations are able to traverse such potential barriers as seen by the non-unity reflection coefficients.

For the bilayer, an estimate of $\lambda_{\rm SPP}$ \cite{roldan13,sensarama10}, using otherwise identical parameters, predicts twice the  SPP wavelength of single layer graphene, thus exceeding the experimental wavelength by a factor of 2.5. 
This indicates an unusually large reduction in the Fermi level of that bilayer segment.

We then determine the carrier mobility from the experimentally obtained damping $\gamma$.
By correcting for the superposition of ohmic damping and radial decay from the excitation point we obtain an ohmic SPP decay constant of 
$\gamma_{p} = 0.12 \pm 0.04$.
Using $\gamma \approx \sigma_1/\sigma_2+\kappa_2/\kappa_1$, with graphene conductivity  $\sigma= \sigma_1+i \sigma_2$ \cite{fei12}, and $\kappa_2/\kappa_1 = 0.04$, this results in a carrier relaxation rate $\Gamma=\frac{\sigma_1 \omega}{\sigma_2} =14 \pm 4$~GHz.
The relaxation rate then relates to the mobility $\mu=e \nu_{\rm F}/\Gamma \hbar k_{\rm F}$ \cite{jablan09}, giving $\mu$~=~1.6$\times$10$^3$~cm$^2$V$^{-1}$s$^{-1}$.

Our derived mobility is in good agreement with previous \emph{s}-SNOM measurements \cite{fei12,chen12}, but lower than typical values for exfoliated samples as determined by transport measurements \cite{geim07}. 
Similarly, our values for $E_F$ and $n$ agree with values derived for samples studied previously by \emph{s}-SNOM \cite{fei12,chen12}, but they and the variations seen due to charge pooling are larger than expected \cite{geim07,castroneto09,martin08,zhang09}.
Recent work has attributed the discrepancy in damping rates and thus mobility to the large SPP frequency and associated wavevectors, which result in incerased in impurity scattering and therefore deviations from assumed DC values \cite{principi13}.
The surprisingly large Fermi energy has not been addressed.

Our results provide for the first time insight into the phase behavior of graphene SPPs. 
In particular it provides complementary information to the amplitude.
It is robust to changes in signal intensity due to the presence of, e.g., bilayer graphene as seen in Fig.~\ref{wedge} or the presence of surface contaminants that otherwise reduce the amplitude signal (see supplement).
The optical phase also provides an additional constraint for our cavity model.
In particular, the absence of independent parameters used to describe the amplitude and phase underscores its validity.
Despite its simplicity, good semi-quantitative agreement is found between the model and the data including the phase relationship between the amplitude and phase.
However, the overall agreement in the amplitude near edges is seen to decrease with dips in the model and spatial shifts in the oscillation extrema.
This can be attributed to changes in electronic structure \cite{fei12} and local field variations near edges \cite{chen12} and underscores the rich complexity of the physical phenomena measurable by this technique.
We note that our parameters do not reproduce the dual peaks in the amplitude near the wedge indicated by a white arrow in Fig.~\ref{wedge}(b).
These can only be reproduced through the use of R~=~1, indicating a local mode as noted previously \cite{chen12}.

We have shown the full amplitude and phase of spatially resolved imaging of graphene SPP's scattered and reflected by discontinuities such as edges, grain boundaries, and defects. 
We describe these results with a simple phenomenological cavity model using no adjustable parameters.
We have observed the effects of charge pooling and associated Fermi level offsets in sample regions with a high density of grain boundaries and folds.
These results highlight the ability of graphene plasmon interferometry and imaging to provide insight into local electronic structure and dynamics in graphene.

{\bf Acknowledgement}
The authors thank Greg Andreev for stimulating this study, as well as Joanna Atkin for fruitful discussions.
This research was supported by the U. S. Department of Energy, Office of Basic Energy Sciences, Division of Materials Sciences and Engineering under Award No DE-FG02-12ER46893.


\end{document}